# A Simple Experimental Model to Investigate Force Range for Membrane Nanotube Formation

*Chai Lor[1], Joseph D. Lopes[2], Michelle K. Mattson-Hoss[3], Jing Xu[2]\* and Linda S. Hirst[2]\**

[1] *Biological Engineering and Small-scale Technologies Graduate Program, School of Engineering, University of California Merced, Merced, CA, USA,* [2] *Physics Department, School of Natural Science, University of California Merced, Merced, CA, USA,* [3] *Developmental and Cell Biology, School of Biological Sciences, University of California Irvine, Irvine, CA, USA*

The presence of membrane tubules in living cells is essential to many biological processes. In cells, one mechanism to form nanosized lipid tubules is *via* molecular motor induced bilayer extraction. In this paper, we describe a simple experimental model to investigate the forces required for lipid tube formation using kinesin motors anchored to 1,2-dioleoyl-sn-glycero-3-phosphocholine (DOPC) vesicles. Previous related studies have used molecular motors actively pulling on the membrane to extract a nanotube. Here, we invert the system geometry; molecular motors are used as static anchors linking DOPC vesicles to a two-dimensional microtubule network and an external flow is introduced to generate nanotubes facilitated by the drag force. We found that a drag force of ≈7 pN was sufficient for tubule extraction for vesicles ranging from 1 to 2 μm in radius. By our method, we found that the force generated by a single molecular motor was sufficient for membrane tubule extraction from a spherical lipid vesicle.

## INTRODUCTION

The lipid membrane is a self-assembled system, which can take on various shapes simply by changing its composition, changing the concentration of specific lipids, or applying a force. Lipids are amphipathic and thus self-assemble into a bilayer in the cell known as the plasma membrane. The membrane forms a barrier that compartmentalizes different cellular processes. Inside the cell, lipid nanotubes are one of various shapes that help facilitate the organization of cellular processes. It has been shown that networks of lipid nanotubes play a role in cellular communication and transport, for example in the immunological synapses (Carlin et al., 2001; McCann et al., 2007), connections between the Golgi and endoplasmic reticulum (Hirschberg et al., 1998; Martinez-Menarguez et al., 1999), and in chemical signaling (Watkins and Salter, 2005). The smooth endoplasmic reticulum is an example of a lipid nanotube system that branches out from the rough endoplasmic reticulum and nuclear envelope. These membrane nanotubes are found to colocalize with fundamental biopolymers in cells known as microtubules; when the microtubules depolymerize, the lipid tubules retract (Terasaki et al., 1986). Recent studies also demonstrated that a minimal system of microtubules and molecular motors were sufficient for tubule extraction (Roux et al., 2002).

Molecular motors are nature's nanomachines, and they drive mechanical translocation of cellular materials along their biopolymer tracks (microtubules) in cells. Kinesin (conventional kinesin or kinesin-1) is a major microtubule-based molecular motor that utilizes the chemical energy stored in ATP to actively step along the microtubule (Schnitzer and Block, 1997). In the absence of ATP, individual kinesin motors associate with and remain strongly bound to the microtubule, with an unbinding energy of ~5 pN per kinesin motor (Uemura et al., 2002). When multiple kinesins are present to bind the same cargo to the microtubule, the unbinding energy (or the force necessary to detach their common cargo) scales linearly with motor number (Mallik et al., 2005; Vershinin et al., 2007, 2008; Shubeita et al., 2008; Jamison et al., 2010; Xu et al., 2012a,b).

In biological membranes, the lipid bilayer behaves like a flexible thin molecular sheet with fluid-like in-plane properties. The lipid molecules forming the bilayer diffuse freely in the plane of the bilayer with a mobility often characterized by the molecular diffusion constant, $D$. The biological membrane has a complicated composition, including many different saturated, unsaturated, and cholesterol lipids plus a large number of protein structures and other membrane associated molecules. Each component in the membrane has an effect on the fluidity and packing structure. For example, the addition of cholesterol into 1,2-dipalmitoyl-sn-glycero-3-phosphocholine (DPPC) membrane decreases membrane fluidity while it increases in DOPC membranes (Mills et al., 2009; Gracia et al., 2010). In addition, membranes composed of mixtures of different lipid components may create in-plane phase separation and domains (or "rafts") with differing elastic properties, for example, in DOPC/DPPC mixtures (Mills et al., 2009; Schmidt et al., 2009). However, in the experiment presented here, we used only one membrane composition with well-known elasticity to simplify our analysis of the multi-component system.

It is widely known that lipid tubules can be extracted from flat bilayers (or vesicles) given enough applied force. Membrane tubules can be extracted from membranes on application of a force with methods, such as micropipette aspiration (Mannerville et al., 2001; Cuvelier et al., 2005), optical tweezers (Cuvelier et al., 2005; Inaba et al., 2005; Koster et al., 2005), molecular motor pulling (Koster et al., 2003; Leduc et al., 2004; Shaklee et al., 2008), and assisted shear flow (Brazhnik et al., 2005; Sekine et al., 2012). Each of these techniques investigates the mechanical forces necessary to deform the lipid membrane into tubules from giant unilamellar vesicles (GUVs) or lamellar structures. Techniques such as force spectroscopy in atomic force microscopy (Scheffer et al., 2001; Picas et al., 2012) and the Langmuir–Blodgett trough (Nichols-Smith et al., 2004) had also been used to quantify the elastic parameters of various membranes without tubulation.

In this study, we coated the glass surface with microtubules for specific binding of kinesin motor proteins, which act as static anchor sites for GUV linkages. By applying a flow to the system, free GUVs were carried by the flow but anchored GUVs experienced a resisting force due to the microtubule-bound motors. Anchored GUVs either pull out lipid nanotubes or remain immobilized or release from the anchor site. This new experimental system is motivated by previous studies where lipid tubulation was induced through a cluster of arriving and departing kinesins walking along the microtubule (Koster et al., 2003). In our case, we anchored the kinesin motors to microtubules and controlled the detachment of kinesin by the force generated on the motor by GUV deformation and displacement.

## MATERIALS AND METHODS

### Materials

All of the lipids 1,2-dipalmitoyl-sn-glycero-3-phosphocholine (DPPC), 1,2-dioleoyl-sn-glycero-3-phosphocholine (DOPC), 1,2-dioleoyl-sn-glybero-3-[(N-(5-amino-1-carboxypentyl) iminodiacetic acid) succinyl] (ammonium salt) (DGS-NTA), and 1,2-dioleoyl-sn-3-phosphoethanolamine-N-(7-nitro-2-1,3-benzoxadiazol-4-yl) (ammonium salt) (DOPE-NBD) were purchased from Avanti Polar Lipids in chloroform and used without further purification. Lyophilized tubulins were purchased from Cytoskeleton (Cat. T240). Recombinant kinesin protein (K560) was purified as previously described (Xu et al., 2012b). Silicon beads of 2.47 μm (SS04N) and 4.56 μm (SS06N) in water were purchased from Bangs Laboratories, Inc.

### Giant Unilamellar Vesicles

To prepare GUVs, we used the method of electroformation to produce GUVs between 1 and 10 μm (Pott et al., 2008). The 94.95 mol% DPPC or DOPC was mixed with 5 mol% DGS-NTA and 0.05 mol% DOPE-NBD to a final concentration of 5 mM in chloroform. DGS-NTA was chosen because it can bind directly by ionic interaction to the HIS-tag of the kinesin protein (Herold et al., 2012). DOPE-NBD is a labeled lipid with absorbance at 460 nm and emission at 535 nm for fluorescence imaging. Droplets of the lipid mixtures were then deposited onto clean indium tin oxide (ITO) slides and allowed to dry in a vacuum chamber overnight to remove excess chloroform. The slides were then assembled into an electroformation chamber to produce GUVs. Two ITO slides were placed between a greased Teflon O-ring with the dried lipid sides facing each other. The chamber was filled with 200 mM sucrose. Once the chamber is assembled, a 2-V sinusoidal voltage with 7.5 Hz was applied across the chamber and incubated at 45°C for 3 h. GUVs are formed by budding from the stack of bilayers that were hydrated at the ITO glass surface. GUV solution were harvested and stored in centrifuge tubes for further use. GUVs produced from this parameter range from 1 to 10 μm.

### Lipid Nanotube Formation

To assemble microtubules, 5 μg/μl tubulin in PEM80 buffer (80 mM PIPES pH 6.9, 2 mM MgCl$_2$, and 0.5 mM EGTA) supplemented with 1 mM GTP and 5% of glycerol was incubated in ice for 3 min and then in water bath at 37°C for 20 min. The assembled microtubules were stabilized by mixing in equal volume in PEM80 buffer supplemented with 50 μM Taxol and 2 mM of GTP and then incubated for another 20 min at 37°C. Flow cells were prepared by sandwiching a poly-L-lysine coated cover glass (22 mm × 40 mm) on top of a microscope glass slide (3 in × 3 in × 1 mm) using two double-sided tape strips. The strips were placed 3 mm apart from each other at the middle of the glass slide creating a flow channel (Figure S1 in Supplementary Material). The resulting flow cell volume was ≈12 μl. To pattern the flow cell with microtubules, 12 μl of microtubule solution (15 μg/ml in PEM80, supplemented with 1 mM GTP and 25 μM Taxol) was introduced to the flow cell. After 5 min of incubation, unbound microtubules were washed away with 12 μl of wash solution (PEM25 supplemented with 1 mM GTP and 25 μM

Taxol). The flow cell was then blocked with 5.55 mg/ml casein supplemented with 1 mM GTP and 25 µM Taxol to prevent non-specific attachment of kinesin onto bare glass. Then, 10 µl of kinesins (4 µM) was introduced into the flow cell to allow for binding to microtubules. After 5 min, excess kinesin was washed away with 12 µl wash solution (PEM80 ed with 1 mM GTP and 25 µM Taxol) and GUVs were introduced into the flow cell and allowed to settle for 10 min. The GUV solution was diluted by 10× with microtubule solution, thus making the GUV interior composed of the sucrose solution. The sucrose makes the GUVs heavy so that they will sink to the surface. To generate lipid tubules, 20 µl wash buffer was drawn through the flow cell using filter paper.

## Lipid Diffusion Measurement Using FRAP

The diffusion constant, $D$, for GUVs containing 5 mol% DGS-NTA was assessed using fluorescence recovery after photo bleaching (FRAP) on supported lipid bilayers. Lipid mixtures of 94.95 mol% DOPC, 5 mol% DGS-NTA, and 0.05 mol% DPPE-NBD (dye) were mixed in chloroform and vacuum dried to remove excess chloroform. Dried lipids were then rehydrated with Milli-pore water to final concentration. Small unilamellar vesicles were formed via tip sonication and drop casted onto a clean glass slide coated with bovine serum albumin. Then the sample was incubated at 50°C for 1 h to allow for fusion on the surface forming a single bilayer. The bilayer was observed using fluorescence microscopy and a small circular region overexposed by high power to bleach the region. The recovery of the bleached region was recorded for analysis of the diffusion constant for the lipid mixture. Snapshots of bleached region recovery were captured by Qimaging camera at 120-s interval for 20 min. Diffusion constant, $D$, was determined by first-fitting Soumpasis formula to the data (Figure S2 in Supplementary Material) (Soumpasis, 1983).

$$f(t) = e^{\frac{-2\tau_D}{t}} \left[ I_0\left(\frac{2\tau_D}{t}\right) + I_1\left(\frac{2\tau_D}{t}\right) \right] \quad (1)$$

where $f(t)$ is the normalized fluorescence, $\tau_D$ is the characteristic time for diffusion, $I_0$ is the fluorescence intensity, and $t$ is the time. $\tau_D$ is determined from the fit, which we can now calculate the $D$ value for a bleached region of radius $\omega$ using

$$\tau_D = \frac{\omega^2}{4D} \quad (2)$$

## RESULTS AND DISCUSSION

Using our experimental system, we observed many examples of successful flow-induced nanotube extraction from lipid GUVs. Kinesin dimers (heads of the kinesin protein) bind to the tubulin of the microtubule. The tail of the kinesin was tagged with histidine polymer, which can bind to the amine group at the head of the DGS-NTA lipid as shown in **Figure 1**.

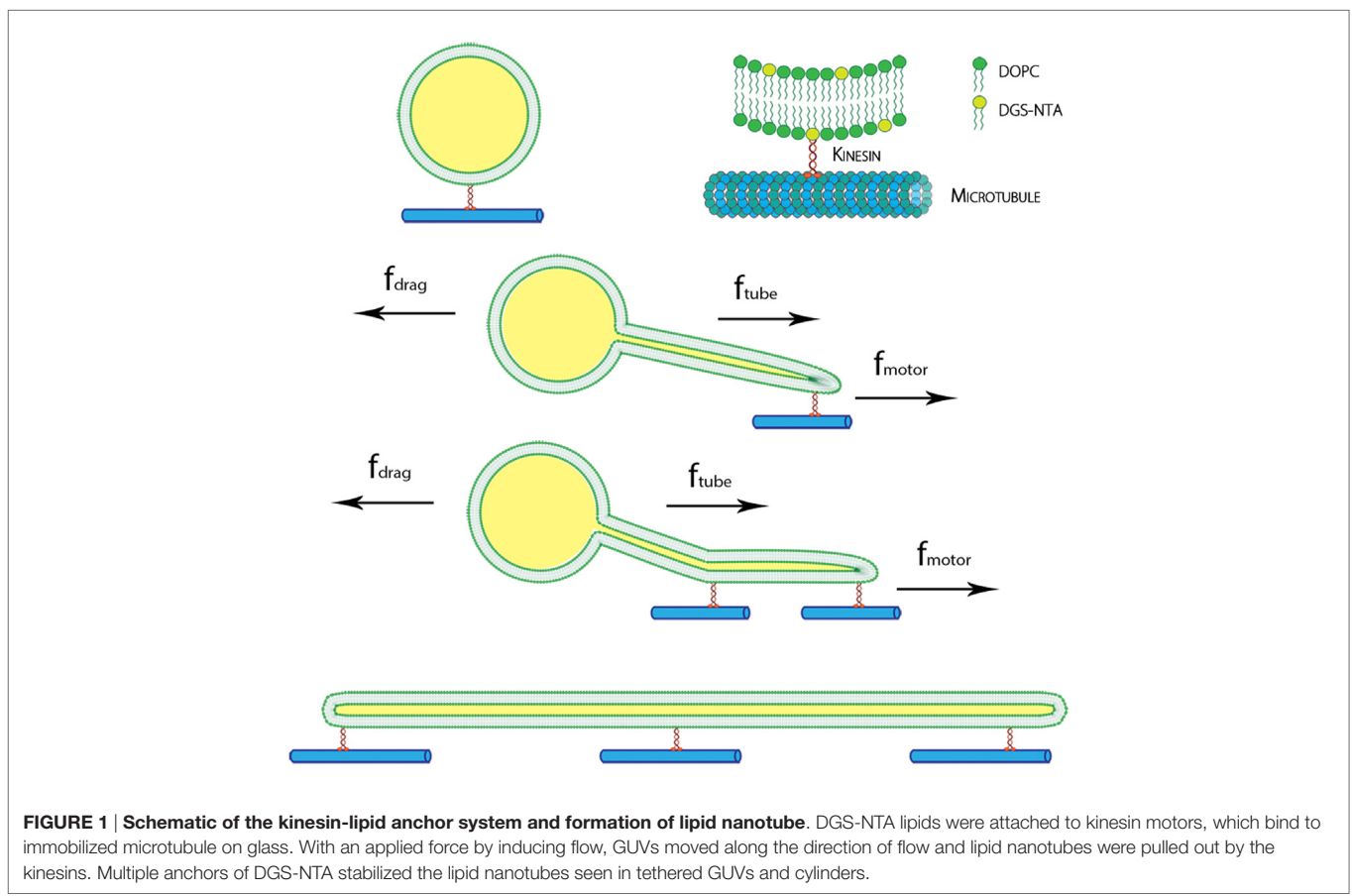

**FIGURE 1 | Schematic of the kinesin-lipid anchor system and formation of lipid nanotube**. DGS-NTA lipids were attached to kinesin motors, which bind to immobilized microtubule on glass. With an applied force by inducing flow, GUVs moved along the direction of flow and lipid nanotubes were pulled out by the kinesins. Multiple anchors of DGS-NTA stabilized the lipid nanotubes seen in tethered GUVs and cylinders.

By using filter paper to induce a flow, the anchored kinesins pulled out nanotubes from the moving GUVs. After a flow was introduced and subsequently stopped, examples of spherical GUVs, tethered GUVs (vesicle with tube pulled out), and fully elongated vesicles which become cylinders were observed as shown in **Figure 2**. Lipid bilayer cylinders as long as 100 μm were generated using this method. A control experiment was done without kinesin, and as expected, all GUVs were washed away from the flow cell without binding to the microtubules.

Under high flow rates, we observed that most GUVs >4.5 μm in diameter were washed away from the flow cell leaving smaller ones that were anchored. However, with a lower flow rate, anchored GUVs with diameters up to 10 μm remain in the flow cell. To quantify the flow rate, we calculated the speed of flowing 2.47 μm glass beads using bright field optical microscopy. For a low flow rate, 20 μl of the wash buffer was deposited at one end of the channel. This created a low flow as the solution spread in every direction due to surface tension. We generated a high flow rate using a strip of 1 cm × 4 cm filter paper to absorb 20 μl of wash solution to the opposite end of the flow cell. We measured an average low flow rate of 61.3 ± 3.7 μm/s and high flow rate of 177.6 ± 18.2 μm/s. We were aware that the flow rate in either case was not constant, and therefore, we calculated an average flow rate for the first 2-s of flow, during which time we expected to generate the fastest flow as the water was wicked into the filter paper. We noted that bigger GUVs were not washed from the flow cell channel under a low flow rate, while the high flow rate was sufficient to detach the big vesicles from the anchor points.

To compare the effects of membrane composition on tabulation, we also prepared GUVs with a different membrane bending rigidity, $k_b$. The bending rigidity is the membrane's resistance to bending deformation. At room temperature, DOPC is in fluid phase ($L_\alpha$) with $k_b$ = 85 pNnm (Rawicz et al., 2000), but DPPC exhibits the gel phase ($L_\beta$) at the same temperature with $k_b$ = 1100 pNnm (Lee et al., 2000). To compare these radically different membrane rigidities in our experiment, we formed GUVs with the DPPC membrane by electroformation and followed up with the above-described procedure to produce lipid nanotubes (Pott et al., 2008). At room temperature, DPPC GUVs are not perfectly spherical but crumpled due to their tilted packing in the bilayer (**Figure 2A**) (Hirst et al., 2013). Therefore, because of their very high bending rigidity, no lipid nanotubes were formed. With DOPC, the bending rigidity is much lower; therefore, it was easier to extract lipid tubules.

**Figure 3** shows a 3-μm diameter GUV being stretched under high flow rate. It was being stretched into an ellipsoid and retracted back to its initial position. There was no lipid nanotube formed. When the flow stopped, the stretched GUV retracted back to a spherical vesicle. A lipid nanotube cannot be formed possibly due to having multiple anchors of DGS-NTA lipids to kinesins. Also, the generated drag force from the size of the GUV may not be high enough to extract a lipid nanotube. A discussion of this phenomenon will be discussed later in the paper.

In **Figure 4**, we see a 2-μm diameter GUV forming a lipid nanotube under high flow rate. This lipid nanotube was formed under a flow of 177.6 ± 18.2 μm/s and stretched to 57 μm in length. Since we imaged the lipid nanotubes using fluorescence microscopy, an optical technique, we cannot accurately measure the radii. However, various methods have been used by other research groups and found that nanotubes extracted from GUVs were between 20 and 110 nm (Roux et al., 2002; Tokarz et al., 2007; Adams et al., 2010; Stepanyants et al., 2012).

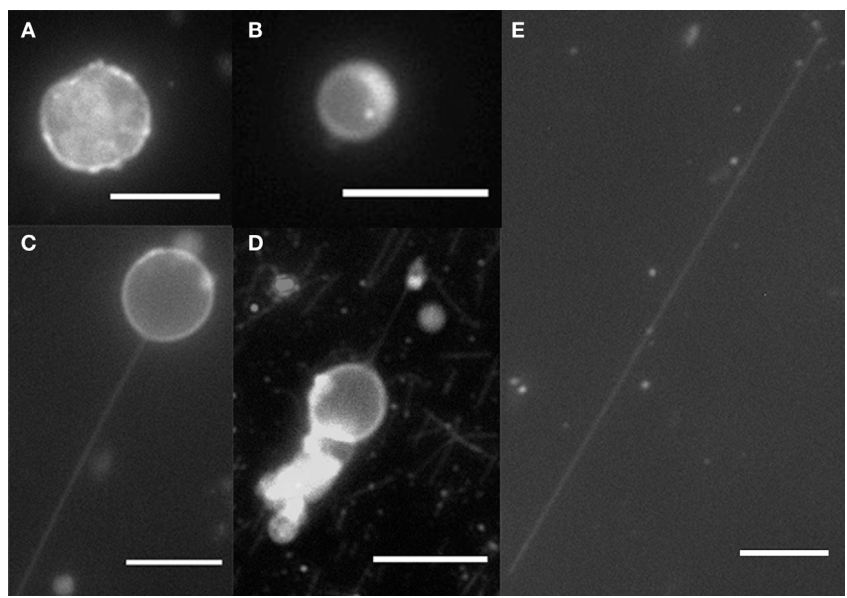

**FIGURE 2 | Fluorescence microscopy of GUVs and nanotubes**. **(A)** DPPC/DGS-NTA (5 mol%) crumple vesicle, **(B)** DOPC/DGS-NTA (5 mol%) vesicle, **(C)** DOPC/DGS-NTA (5 mol%) tethered GUV, **(D)** DOPC/DGS-NTA (5 mol%) tethered GUV with labeled microtubules, and **(E)** DOPC/DGS-NTA (5 mol%) lipid nanotube. Scale bar: 10 μm.

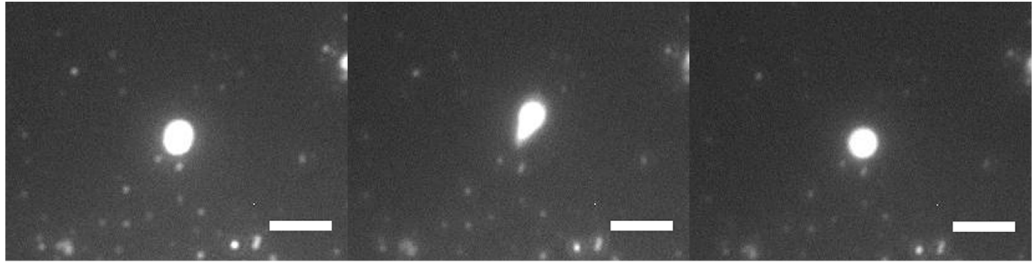

**FIGURE 3 | Fluorescence microscopy of GUV retraction under high flow**. A DOPC/DGS-NTA (5 mol%) GUV of 3 μm in diameter was being stretched by the flow with an anchor. The GUV stretched into an ellipsoid and retracted back into GUV. Scale bar: 5 μm.

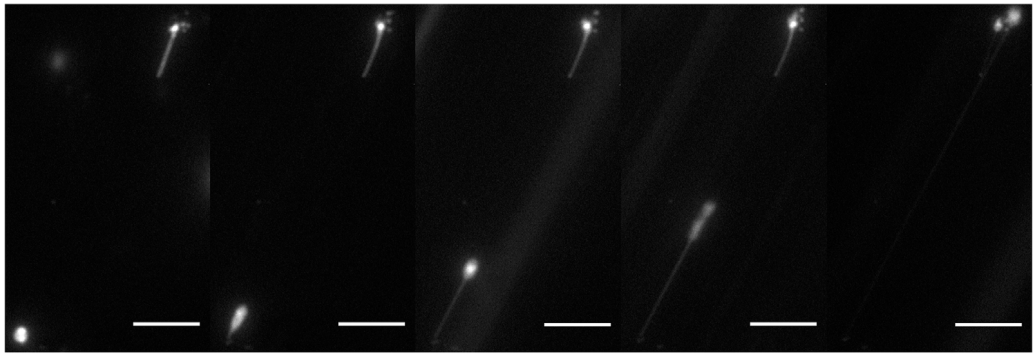

**FIGURE 4 | Fluorescence microscopy of lipid tubulation under high flow**. A lipid nanotube was extracted from a DOPC/DGS-NTA (5 mol%) GUV of 2 μm in diameter by the applied flow. The nanotube was about 57 μm in total. Scale bar: 10 μm.

## Discussion

The lipid membrane forms a bilayer in aqueous solution and can be represented as a soft flexible elastic sheet that can bend out of plane or stretch laterally by increasing the distance between each lipid molecule. Applying a point force perpendicular to the bilayer will deform the membrane from its equilibrium curvature potentially causing a tube to be extracted. For our system, a GUV binds to a kinesin that anchors to a microtubule. When a flow is applied, a nanotube is extracted from the moving GUV due to the force applied by the anchored kinesin. In a flow, GUVs experience a drag force given by Stoke's equation:

$$f = 6\pi\eta r v \quad (3)$$

where $r$ is the GUV radius under a fluid flow of velocity, $v$, in a medium of viscosity, $\eta$. The viscosity of a dilute aqueous solution near the glass surface in the flow cell can be estimated as $\eta = g$ ($1e^{-3}$ N s/m²), where $g$ is a correction factor for flow close to a surface according to Faxen's law[1].

$$g = \left(1 - \frac{9}{16}\left(\frac{r}{h}\right) + \frac{1}{8}\left(\frac{r}{h}\right)^3 - \frac{45}{256}\left(\frac{r}{h}\right)^4 - \frac{1}{16}\left(\frac{r}{h}\right)^5\right)^{-1} \quad (4)$$

[1] http://lben.epfl.ch/files/content/sites/lben/files/users/179705/Optical%20Trapping%20Handout.pdf

where $h$ is the distance from the glass surface and $r$ is the GUV radius. For example, in our flow cell, we approximate $g = 2.36$ for a 1-μm radius GUV, 100 nm above the surface (Hirokawa and Noda, 2008). The $g$ value is typically between 1 and 3 and only valid for when $h - r \geq 0.02r$. We calculated the drag force shown in **Figure 5** for DOPC GUVs under high and low flow rates.

We can first considered a simple non-deformable system of either glass beads or GUVs with a high bending rigidity, such as DPPC in the gel phase, anchored to a microtubule by kinesin. In such a case, we do not expect a nanotube to form, and we can write that $f_{drag} \leq f_{motor}$. If $f_{drag}$ (drag force on the GUV) exceeds the maximum attachment force of the cumulative motors, $f_{motor}$, then the anchored kinesin/GUV should detach from the microtubule. If $f_{drag}$ is less or equal to the $f_{motor}$, then the GUV is immobilized. We see clear evidence of this effect with DPPC GUVs, they either detach and are washed away by the flow or remained attached without tube formation.

In the case of softer elastic membranes, such as DOPC in the fluid phase at room temperature, lipid nanotubes can be pulled from the GUV as the membrane deforms at the kinesin anchor point. We can describe the free energy cost of extracting a tubule of radius $r$ and length $L$ from the GUV using the Helfrich formula assuming no increase in membrane surface area $A$ and volume $V$ (Deuling and Helfrich, 1976; Derenyi et al., 2002).

$$F = \left(\int \frac{k_b}{2}(2H)^2 dA\right) + \sigma A - \rho V - fL \quad (5)$$

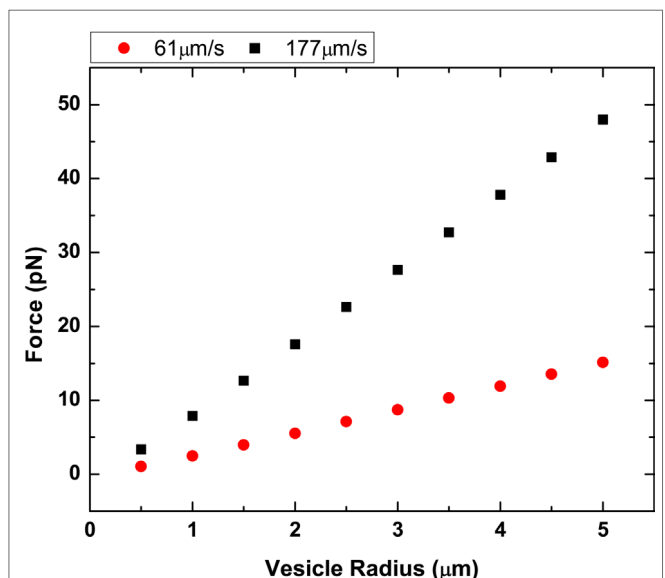

FIGURE 5 | **Plot of vesicle radius vs. drag force**. Two different flow rates at 61.3 and 177.6 µm/s were plotted using both Stoke's law and Faxen's law.

where $k_b$ is the membrane bending rigidity, $H$ is the mean curvature, $(1/R1 + 1/R2)/2$, $\sigma$ is the surface tension, and $\rho$ is pressure. In tethered GUVs, the energy of surface tension (minimizing the surface area) acts against the energetic costs of bending, pressure induced volume increase, and tube elongation to minimize the total free energy. Without the action of deformation forces, the bilayer takes the shape of a spherical vesicle. In the formation of the nanotube, the pressure difference inside the GUV and tether is 0, otherwise it would rupture. In order to find the force required to extract a tube of radius $r$, we calculated $\partial F/\partial R = 0$ and $\partial F/\partial L = 0$ for an extracted tube in equilibrium, thus arriving at

$$r = \sqrt{\frac{k}{2\sigma}}, f = 2\pi\sqrt{2\sigma k} \qquad (6)$$

which leads to a single equation, $f = 2\pi k/r$. From here, we can calculate the force, $f_{tube}$, it takes to pull a tube of radius $r$ out of a GUV with bending rigidity, $k_b$. In our system, the $k_b$ values for DOPC and DPPC are 85 pNnm (Rawicz et al., 2000; Pan et al., 2008) and 1100 pNnm (Lee et al., 2000), respectively. We use fluorescence microscopy to observe the lipid nanotubes; however, measuring the tube radius is quite complicated due to the limits of optical resolution. Several prior studies have determined the radius of these tubes to be in the range of 20–110 nm (Roux et al., 2002; Tokarz et al., 2007; Adams et al., 2010; Stepanyants et al., 2012). If a tube has a radius of 50 or 100 nm, then we get tube forces ($f_{tube}$) of 10.7 and 5.35 pN, respectively, using our analysis. The difference between the $f_{drag}$ and the sum of $f_{tube}$ for all attached motors results in the net force exerted by the motors on the vesicle, $f_{drag} - n f_{tube} = f_{vnet}$, where $n$ is the number of tubes generated. Although this equation takes into consideration the possibility of multiple tubes, such a state is less likely and coalescence tether experiments have shown that nanotubes generated close to each other will fuse into a single nanotube (Cuvelier et al., 2005).

We observed no more than one lipid nanotube per GUV in our experiments.

In our experiments, we observe three different cases:

(1) $f_{vnet} > n f_{motor}$: large GUVs are washed from the flow cell after the onset of flow. Large GUVs generate a $f_{vnet}$ higher than the $n f_{motor}$ from anchored kinesin(s). Kinesin motors' resisting force is not sufficient to overcome the strong drag force, and thus, the large GUV is swept away from the surface.
(2) $f_{vnet} \leq n f_{motor}$: under our high flow rate of ≈177 µm/s, smaller GUVs <5 µm anchored to the microtubule *via* kinesin are not washed from the flow cell. $f_{vnet}$ is not strong enough to detach the motors from their bound state. Depending on the number of motors being attached to the microtubule and their distribution, a tube may be pulled out from the vesicle, given a sufficiently deformable membrane.
(3) $f_{vnet} \ll n f_{motor}$: the net force from the drag force and tube force is too low compared to the max force from the total number of bound kinesins. In this group of GUVs, no nanotube is generated because it would cost too much energy to make tubes per anchored motor. In this case, having too many bound anchor sites and a stiff membrane prohibits the GUV from either washing away or pulling out nanotubes.

In **Figure 6**, we plotted two phase diagrams indicating the theoretical result (nanotube formation, GUV detachment, or immobilization) as a function of number of attached motors and vesicle radius for the two different flow regimes. We calculated the theoretical minimal number of motors needed to successfully anchor a GUV and generate a nanotube given the calculated drag force (**Figure 5**). As described in case 1, large GUVs were washed from the flow cell under high flow – they generated a high drag force that overcame the resistance force from the combined anchored motors. This calculation of motor number assumed that each motor has a maximum binding force of 5 pN to the microtubule. Here, we assumed that lipid nanotubes extracted from the GUVs are 67.6 nm in radius. This value was an estimation based on $f = 2\pi k/r$ for the smallest GUV in which nanotubes were observed. These calculations demonstrated that nanotubulation requires a minimum of just one or two motors under both flow conditions. According to the graph, if there are more motor/lipid linkages than the estimated value, then the GUV does not form a nanotube and remains immobilized. However, if there are fewer kinesin/DGS-NTA linkages, then the GUV flows away.

A theoretical estimation of the number of motors required for nanotube formation is shown in **Figure 6**, and our experiments showed a general qualitative agreement with these results. However, we did not observe nanotubes over the full range indicated in the phase diagrams. In **Figure 6B**, we plotted for GUVs up to 5 µm in radius but only observed nanotubes for GUVs between 1 and 2 µm. We believed it is possible to form nanotubes in GUVs >2 µm by increasing the concentration of kinesin motors and microtubules on the substrate. However, these linkages have to be distanced enough from each other, which can be quite difficult. If too many motors anchored are close together, which is likely the case especially on the same microtubule, only the motor at the tip experiences the force from the vesicle. Having too many anchor sites may not cause a point deformation but rather an area deformation, stretching the GUVs into an ellipsoid. An example of this was seen in **Figure 3**. We observed GUVs stretched into ellipsoids in the presence of the flow and retracted back to a sphere when the flow was stopped.

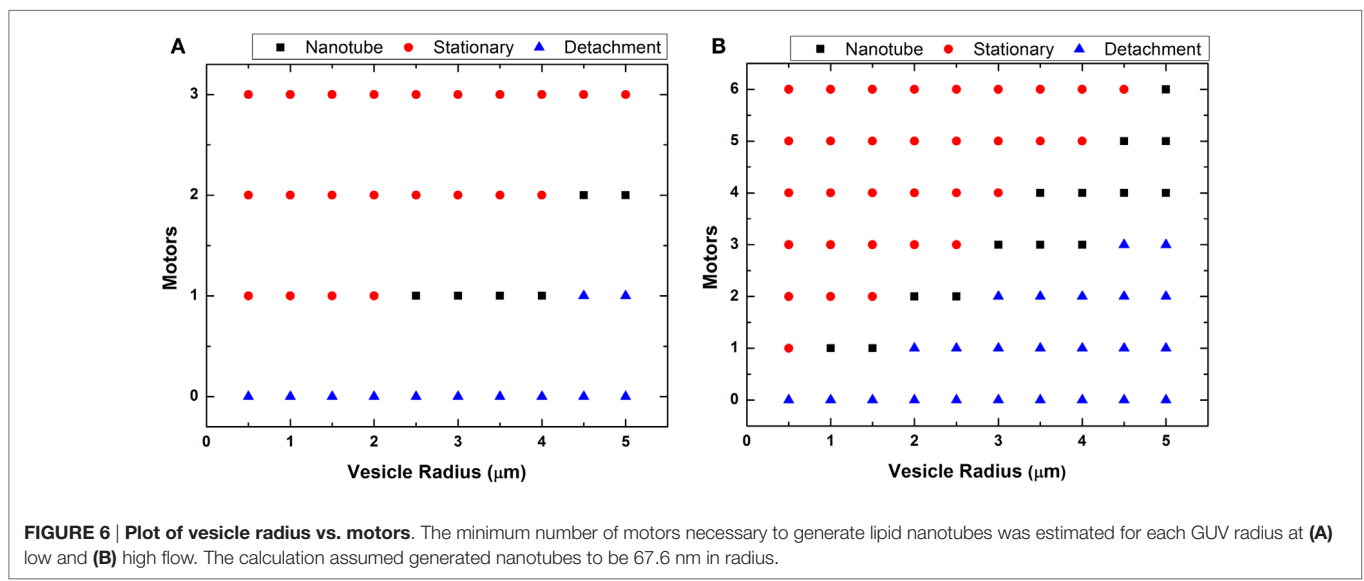

**FIGURE 6** | **Plot of vesicle radius vs. motors**. The minimum number of motors necessary to generate lipid nanotubes was estimated for each GUV radius at **(A)** low and **(B)** high flow. The calculation assumed generated nanotubes to be 67.6 nm in radius.

Tethered GUVs and fully stretched GUVs forming cylinders are evidence of multiple linkages. During the flow process, tubulation can be observed from displaced anchored GUVs. When the flow slowed down or stopped, GUVs retracted back to their initial location. However, during the process of tubulation, if a GUV is able to make another lipid/kinesin linkage, then a "tethered" GUV is stabilized because the tube only retracts back to its most recent anchor. For fully stretched vesicles (cylindrical tubes), both ends of the cylinder must have anchor sites. The probability of making additional tube-microtubule motor linkages is high with our lipid mixture of DOPC with 5 mol% of DGS-NTA. The lateral diffusion of DOPC/DGS-NTA lipid molecules at room temperature is high, $9.01 \pm 0.58$ $\mu m^2/s$, compared to pure DOPC at 8.2 $\mu m^2/s$ (Figure S2 in Supplementary Material) (Lindblom and Oradd, 2009). The hydrophobic tails of both lipid molecules are the same, but DGS-NTA have a bigger hydrophilic head, which may contribute to increased lateral diffusion. This suggests that for a moving GUV traveling along microtubule coated with kinesin motors, there is a high probability that a DGS-NTA lipid will be present at the bottom of the GUV to bind with incoming kinesin. Increasing the concentration of kinesin motors potentially increases the number of fully stretched GUVs as extra-binding sites can act to hold the tubule in place; however, there will be a trade-off because too many motors present in the system will affect its ability to even form a nanotube and elongate in the case of multiple initial bindings. The fully stretched tethered tubules we observed (**Figure 2E**) were very stable and straight and they all stretched in the direction of the flow. We also observed that once stabilized, further stretching due to additional flow did not occur. This was likely because with a cross-section of only $\approx$140 nm, the nanotubes did not generate enough drag force to detach the multiple-bound kinesin motors from the microtubules and allow further extension.

## CONCLUSION

We have developed a simple experimental system to generate flow-induced lipid nanotubes using kinesin molecular motors anchored to microtubules. By this method, we demonstrated that one can estimate the number of motors necessary for anchoring and tubulation. Our calculations, confirmed by experiments, demonstrated that it takes as little as one to two motors to pull out a tube from a GUV provided we are in a suitable GUV size range. The fabrication of the flow cell and experimental procedure is simple, quick, and should provide fast readout of the motor count if tubes are observed. A flow cell with better flow rate control will be fabricated in the future for quantitative analysis. The presence of nanotube stability requires multiple anchors for GUVs that do not fully retract back to its initial position and fully stretched nanotubes. This may provide possible insight to stability of nanotube networks in cells.

## AUTHOR CONTRIBUTIONS

CL and JL carried out the experiments and analyzed the data; MM-H purified proteins; LH, JX, and CL designed the experiments and wrote the paper.

## FUNDING

This work was supported by the National Science Foundation (DMR BMAT 0852791 to LH), the UC Merced Health Sciences Research Institute Biomedical Seed Grant (to LH and JX), and the UC Merced Academic Senate Committee on Research Award (to JX and LH).

## SUPPLEMENTARY MATERIAL

The Supplementary Material for this article can be found online at http://journal.frontiersin.org/article/10.3389/fmats.2016.00006